\documentclass[a4paper]{article} 
\usepackage{setspace}
\usepackage{fancyhdr}
\usepackage[utf8]{inputenc}

\usepackage[T1]{fontenc}
\usepackage{textcomp}
\usepackage{graphicx}
\usepackage{mathtools}
\usepackage[normalem]{ulem}
\usepackage{hyperref}
\usepackage[table,xcdraw]{xcolor}
\usepackage{amsmath}
\usepackage{subcaption}
\usepackage{csquotes, lmodern}
\usepackage{soul, todonotes}
\usepackage{authblk}
\usepackage[margin=1.0in]{geometry}
\usepackage[backend=biber, sorting=none, style=nature]{biblatex}

\author[1, 2, 4]{Gy\"{o}rgy K\'{a}lvin}
\author[1]{P\'{e}ter Vancs\'{o}}
\author[1]{M\'arton Szendr\H{o}}
\author[1, 3, 4]{Konr\'{a}d Kandrai}
\author[1, 4]{Andr\'{a}s P\'{a}link\'{a}s}
\author[1, 3]{Levente Tapaszt\'{o}}
\author[1, 4 *]{P\'{e}ter Nemes-Incze}

\affil[1]{Hungarian Research Network, HUN-REN Centre for Energy Research, Institute of Technical Physics and Materials Science, 1121 Budapest, Hungary}
\affil[2]{ELTE, Eötvös Loránd University, 1117 Budapest, Hungary}
\affil[3]{Budapest University of Technology and Economics, 1111 Budapest, Hungary}
\affil[4]{MTA - HUN-REN EK Lendület "Momentum" Topology in Nanomaterials Research Group}
\affil[*]{\small \emph{corresponding author, email: nemes.incze.peter@ek.hun-ren.hu}}

\title{Revealing the impact of ambient molecular contamination on scanning tunneling microscopy and spectroscopy of layered materials}

\begin{document}

\maketitle



\pagenumbering{arabic}

\section*{Abstract}
Hydrocarbon contamination is an ever-present factor to consider in surface science measurements.
In the case of van der Waals material surfaces, the structure of this contamination has become known in recent years as a self-assembled layer of normal-alkanes, resulting from a few days' exposure to ambient air.
Knowledge of its composition and structure enables systematic investigation of its influence on surface properties.
Here, we investigate the effect of this contamination on scanning tunneling microscopy (STM) and spectroscopy measurements by comparing clean and ambient alkane-contaminated surfaces of graphite.
Our results reveal that the ambient alkane layer suppresses the well-known phonon-induced gap near the Fermi energy, resolving a long-standing inconsistency in STM studies, where this feature is often absent.
Furthermore, we show that the presence of the contamination layer alters the current-distance ($I(z)$) characteristics, flattening its exponential decay by a factor of 1.5 to 5 compared to the clean surface.
This change arises from extra conductance channels through the alkane layer alongside the tunnel junction, as the tip penetrates the contaminant overlayer.
Finally, based on the $I(z)$ characteristics, we provide a practical guide to detect the presence of surface contamination in STM measurements.

\section*{Introduction}
Hydrocarbons~\cite{palinkas,Li2013-ci,Gallagher2016-io,Tilmann2023-ni,Martinez-Martin2013-wp,Temiryazev2019-dc,Thakkar2022-ww} and water~\cite{Uhlig2021-wn} are universally present on surfaces, including van der Waals materials.
After cleaving the surface, they appear within minutes~\cite{palinkas} and can not be completely removed even in a glove-box environment, leading to bubble formation in heterostructures, that can only be eliminated by exfoliating in ultrahigh vacuum~\cite{Wang2023-vv}.
The exact structure of the initial contamination, right after exfoliation, is unknown, but over a few days any adsorbates are replaced by a self-assembled monolayer of normal-alkanes (n-alkanes)~\cite{palinkas}.
These molecules are common in the environment and are universally present on 2D van der Waals materials~\cite{palinkas,Tilmann2023-ni,Gallagher2016-io}.
This surprising discovery, that mid-length n-alkanes (19-26 carbon atoms) can spontaneously form a contaminant monolayer on vdW materials under ambient conditions~\cite{palinkas} leads to the question, how do they affect the scanning tunneling microscopy (STM) measurements on these systems?

N-alkanes are intrinsically large gap ($>8$ eV) insulators~\cite{Johnson2022-yt}.
The presence of such molecules on a metallic surface can reduce the tunneling barrier height, via well known mechanisms, such as the "pillow effect"~\cite{Vzquez2007} or an induced dipole moment.
Here we show that the STM tip nearly always penetrates the molecular layer, leading to anomalies in the current distance curves and tunneling spectroscopy, due to contact between the tip and the surface molecular layer.
The conductance through an alkane chain is in the regime of the tunneling conductance, of the order of $1$ nS~\cite{transport}.
In STM break junction measurements, Liu et al.~\cite{transport} showed that the conductance of dicarboxylic acid-terminated alkanes decreases exponentially as we increase the length of the chain and the attenuation factor is $0.69$ per methyl unit.
Thus, alkane molecules between an STM tip and sample are capable of making single or multi-molecule conductive junctions which form a parallel resistance to the tunnel junction, altering the measured tunneling current.
Using low temperature (8.7 K) STM measurements, we investigate how the presence of the n-alkane molecules influences the tunneling current, when the tip penetrates the molecular layer.
We show that this leads to an anomalously small decay constant in the tunneling current ($I$) as the tip - sample distance ($z$) is increased.
Since the regime where the tip is measuring above the contaminant layer is unavailable via $I(z)$ measurements, we explore the intrinsic decay rate of the surface wave functions by \emph{ab-initio} calculations.
Finally, we show that the tunneling differential conductance measurements ($dI/dV$) are mostly unaffected since the alkanes have large HOMO-LUMO gaps well in excess of the bias voltages applied during tunneling.
Thus the molecules do not introduce features in the $dI/dV$ spectra~\cite{palinkas}, with one notable exception: alkane contamination suppresses the phonon gap~\cite{phonongap} in graphene and graphite.
This solves the long standing puzzle of why the phonon-induced gap~\cite{phonongap} is clearly visible in some STM measurements~\cite{Sun2023-jt,Zhang2008-hf,Balgley2022-fy,Telychko2022-yk,Natterer2015-xy,Seifert2024-kd}, while not being present in others~\cite{Andrei2012,Kanasaki2009}.


\section*{Experimental Details}

To examine how airborne n-alkane contamination affects STM and scanning tunneling spectroscopy measurements, we prepared two sets of samples.
For a clean control surface, we exfoliated highly ordered pyrolytic graphite (HOPG) under ultra-high vacuum conditions, at a pressure below $5 \times 10^{-11}$ Torr.
These samples were immediately transferred to the STM chamber for measurement at a temperature of 8.7 K.
To assess the impact of alkane contamination, we exposed freshly cleaved HOPG samples to ambient air for two weeks~\cite{palinkas}, with samples stored in either a plastic Petri dish (VWR International, LLC) or a plastic box (Ted Pella Inc.).
We verified the presence of n-alkanes using atomic force microscopy, following the method described in our earlier work~\cite{palinkas}.
Hereafter, we refer to these samples as "contaminated".
After AFM characterization the contaminated samples were loaded into the STM vacuum chamber for measurements at low temperature.
As for the STM tip, we calibrated mechanically cut Pt/Ir tips on a clean Au(111) surface~\cite{tipfabr}.
Repeated pulses and indentation of the sample surface modify the STM tip, and the process is repeated until the herringbone reconstruction and the Au(111) surface state are clearly visible in the dI/dV spectrum without any anomalous feature.
Furthermore, measuring the tunneling current ($I$) as a function of tip sample separation $z$ should show an exponential decay of the current.

All STM and scanning tunneling spectroscopy measurements were performed using an RHK PanScan Freedom system operating at $T = 8.7$~K in UHV conditions with pressure better than $5 \times 10^{-11}$~Torr.
The STM tips were fabricated by mechanically cutting Pt/Ir wire (90\%/10\%).
Differential conductance $(dI/dV)$ spectra were obtained using a lock-in amplifier with a small AC modulation added to the bias voltage: 10 mV at a frequency of 1267 Hz.
Topographic imaging was performed with a sample bias of $0.5$~V and various tunneling current setpoints.
All spectra shown in the main text represent raw data without any smoothing

\section*{Computational Methods}

First-principles calculations were carried out within the framework of density functional theory (DFT) as implemented in the \textsc{VASP} package~\cite{Kresse1999-tr}.
The plane-wave basis set and projector augmented-wave (PAW) method~\cite{Blochl1994-ya} were employed.
The exchange-correlation functional was treated within the generalized gradient approximation (GGA) using the Perdew--Burke--Ernzerhof (PBE) parametrization~\cite{Perdew1996-kb}, and long-range van der Waals interactions were included through the DFT-D2 method of Grimme~\cite{Grimme2006-we}.
The computational model consisted of rectangular supercells containing AB-stacked graphite and C\textsubscript{17} alkane molecules, with lattice constants $a = 31.98$~\AA\ and $b = 12.78$~\AA.
A vacuum region of $40$~\AA\ was included along the $z$-direction to eliminate spurious interactions between periodic images.
Atomic positions were relaxed using the conjugate-gradient algorithm until all forces were smaller than $0.01$~eV/\AA.
The plane-wave cutoff energy was set to $600$~eV.
The Brillouin zone was sampled using a $\Gamma$-centered $2 \times 6 \times 1$ Monkhorst--Pack $k$-point mesh.

\section*{Results and Discussion}

\begin{figure}[h!]
    \centering
    \includegraphics[width=1 \textwidth]{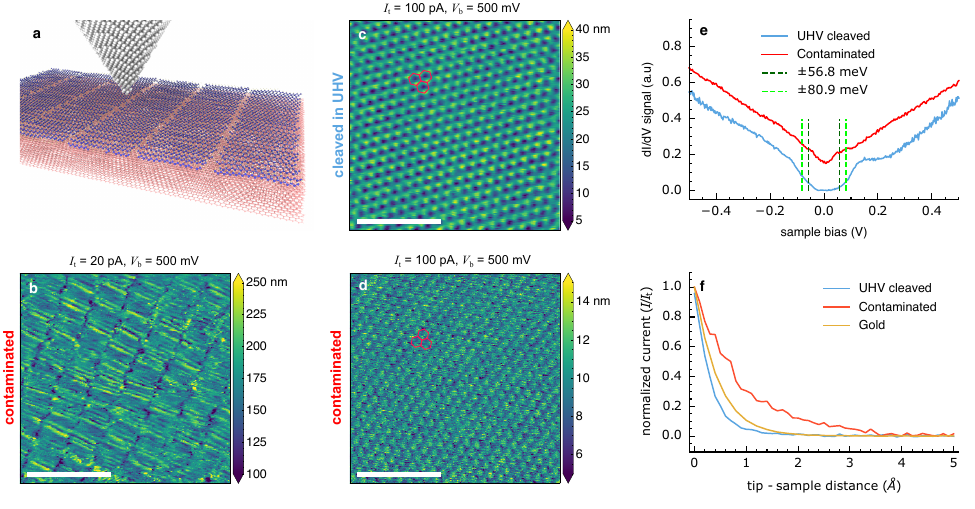}
    \caption{\textbf{The effect of ambient contamination in STM measurements of graphite.}
    \textbf{(a)} Schematic image of the n-alkane self-assembled monolayer stemming from ambient air contamination~\cite{palinkas} on a graphite surface.
    While scanning, the STM tip can either penetrate this layer, or can scan above it.
    \textbf{(b)} Constant current STM topography image of contaminated graphite, with the STM scanning above the contaminant layer.
    The self-assembled structure of the n-alkanes is visible~\cite{palinkas}.
    \textbf{(c)} Constant current STM topography image of UHV cleaved graphite sample.
    \textbf{(d)} Constant current STM topography image when the tip penetrates the alkane layer.
    The honeycomb lattice of graphene is marked with red hexagons.
    The images represent raw data, processed only by line-by-line linear background subtraction along the fast scan direction.
    Scale bars: 5 nm in (b), 2 nm in (c, d).
    Tunneling current setpoint ($I_{\mathrm{t}}$) and sample bias ($V_{\mathrm{b}}$) values are displayed on the top of the panels.
    Data in (b) and (d) are measured with the same tip.
    \textbf{(e)} Representative tunneling conductance measurement ($dI/dV$) performed on UHV cleaved and contaminated graphite samples.
    The green dashed lines show the position of the phonon induced gap edges~\cite{Natterer2015-xy}.
    Spectra are offset for clarity.
    Lock-in parameters: 10 mV modulation at 1267 Hz.
    \textbf{(f)} Single examples of $I(z)$ spectroscopy on clean, contaminated graphite and gold samples, by measuring the tunneling current as the tip is retracted.
    The current of the curves are normalized to 1, by dividing by the setpoint current ($I_{\mathrm{t}}$), for better comparison of the decay.
    The tip - sample distance defined by the current setpoint is chosen as $z = 0$.
    }
    \label{fig:1}
\end{figure}

The self-assembled monolayer of the n-alkanes is schematically represented in Fig.~\ref{fig:1}a, the structure of which has been systematically explored in our previous work~\cite{palinkas}.
All the results presented below are reproduced using multiple tips and on different areas of the samples, as well as on thermally deposited n-alkane layers (dotriacontane C$_{32}$H$_{66}$), for more details see supplementary information.
The contamination layer can be imaged at low temperature, at atomic resolution, if the STM tip is far enough away from the surface~\cite{palinkas}, meaning low tunneling current setpoints need to be used, such as $I_{\mathrm{t}}$ = 20 pA in the example shown in Fig.~\ref{fig:1}b.
If the tip does not perturb the molecular layer, the atomic resolution image of the alkanes is dominated by the $\pi$ electrons of the graphite substrate, with the outer H atom on every second methyl group along the alkane backbone having the highest apparent height in STM topography~\cite{palinkas}.
The exact apparent height and thus topography contrast is determined by the local rotation of the molecules and position of the alkane molecule with respect to the graphite substrate~\cite{palinkas}.

Increasing the setpoint to 50 pA and above ($V_{\mathrm{b}} = 50$~mV), the self-assembled monolayer becomes undetectable, because the STM tip penetrates through the alkane layer.
When the STM tip perturbs the molecular monolayer one can observe the atomic structure of the graphite/graphene underneath.
In this case the atomic resolution image becomes more blurry and noisy.
For a characteristic example of this, see Fig.~\ref{fig:1}d, where we present raw topography data for the contaminated surface and the one cleaved in UHV (Fig.~\ref{fig:1}c).
The difference is quite striking especially if we compare the contaminated atomic resolution image to the raw image of the clean surface.
The increased noise arises because, in addition to tunneling between the STM tip and the graphite sample, charge transport also occurs through the alkane molecules surrounding the tip.
Evidence for this comes from STM break junction experiments, on n-alkanes, for example by Liu et al.~\cite{transport}.
Considering that the conductivity of an alkane molecule in a break junction experiment decays exponentially with the number of carbon atoms in the chain~\cite{transport,van-Veen2022-fq}, we can get an estimate of the conductance contribution of the alkane contaminant.
Using the decay constant from the paper by Liu et al~\cite{transport}, we list in Table~\ref{table1:label} the conductance of single n-alkane molecules contacted by metallic leads as a function of the number of CH$_2$ groups.

\begin{table}[h!]
    \begin{center}
    \small
        \begin{tabular}{|*{10}{c|}}
        \hline
        Number of CH$_2$ groups & \textbf{5} & \textbf{6} & \textbf{7} & \textbf{8} & \textbf{9} & \textbf{10} & \textbf{11} & \textbf{12} & \textbf{13} \\
        \hline
        Estimated conductance (pS) & 9070.07 & 2281.83 & 574.06 & 144.42 & 36.33 & 9.14 & 2.30 & 0.58 & 0.15 \\
        \hline
        \end{tabular}
    \caption{Table of estimated conductance values of alkane chains. Based on the decay constant measured by Liu et al~\cite{transport}.}
    \label{table1:label}
    \end{center}
\end{table}

\begin{figure}[h!]
    \centering
    \includegraphics[width=1\textwidth]{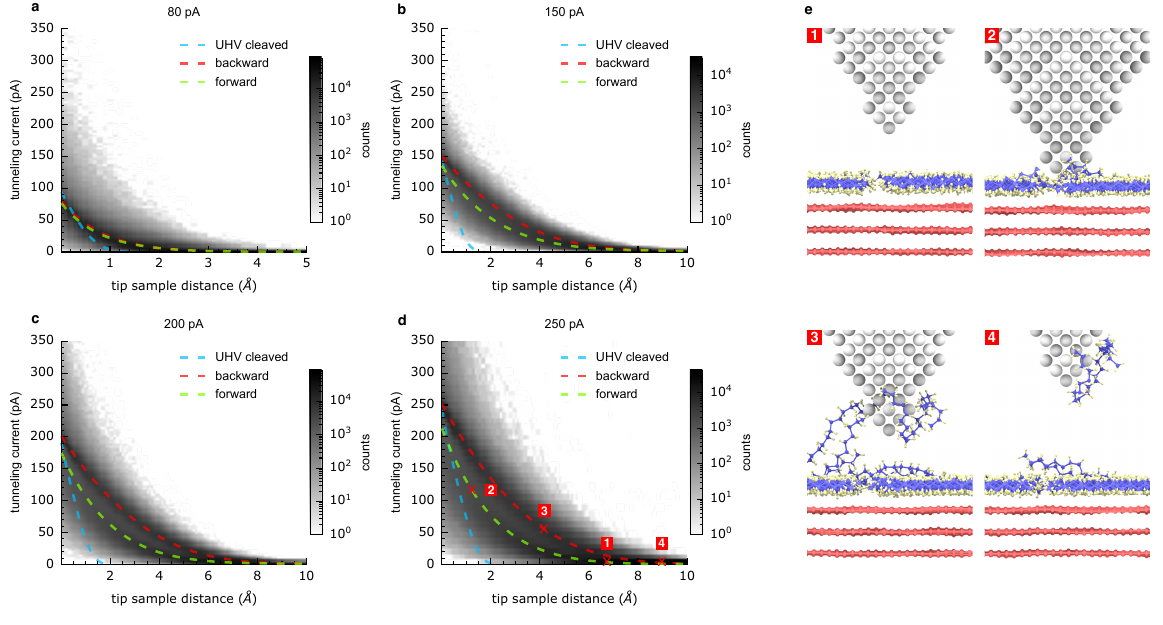}
    \caption{
    \textbf{Influence of surface contamination on $I(z)$ spectra.} \textbf{(a-d)} $I(z)$ spectroscopy curves presented as 2D histograms measured on a contaminated HOPG sample with increasing setpoint.
    The histogram color scale presents the number of data points in a specific tip - sample distance ($z$) and tunneling current ($I$) range.
    The histograms are composed of data from more than 40000 individual $I(z)$ spectra, measured by first retracting the tip from the surface ("backward"), then approaching the sample ("forward").
    The red and green dotted curves represent the averaged forward and backward curves.
    The blue curves are measured on UHV-cleaved samples.
    Setpoint $I_{\mathrm{t}}$ values are shown on top, $V_{\mathrm{b}} = 500$~mV.
    All spectra were recorded with the same tip.
    \textbf{(e)} Schematic representation of the tip-sample interaction with the n-alkane layer, for different tip-sample separations as marked by numbers in (d).
    In all panels, the tip - sample distance defined by the current setpoint is chosen as the $z = 0$ value.
    }  
    \label{fig:2}
\end{figure}

As an order of magnitude, the conductance of an alkane molecule of $9$ to $7$ carbon atoms is the same order of magnitude as the tunneling conductance of $200$ pS in the case of the STM image in Fig.~\ref{fig:1}d.
When the tip is scanning within the molecular layer, numerous alkane molecules make contact between the STM tip and the sample, see a schematic representation of this in Fig.~\ref{fig:2}e.
Furthermore, the contact through a single molecule is most probably not along the ends of the 19-26 CH$_2$ group long chain~\cite{palinkas}, but also through any position along the alkane backbone.
Thus, as the tip scans, the constantly detaching and reforming tip - alkane contact can induce a noise~\cite{Kockmann2009} with fluctuations up to the order of the magnitude of the tunneling conductance and beyond.
This makes the topography images much more noisy, when the tip is scanning within the alkane layer, as shown by the comparison in Fig~\ref{fig:1}c,d.
The inevitable rupture and creation of molecular conduction channels is also evident in the significantly larger current fluctuations of current ($I$) - distance ($z$) curves, as discussed later.
In Fig.~\ref{fig:1}f, we compare individual $I(z)$ curves measured on a contaminated surface, a clean surface cleaved in UHV, and a gold (111) surface during tip preparation.
The increased current noise observed on the contaminated surface is evident.

Next, we investigate the impact of alkane contamination on the $dI/dV$ spectra.
Since the molecular layer exhibits a large HOMO-LUMO gap, functioning as a large gap insulator, no molecule-specific peaks or features are observed in the $dI/dV$ spectra.
Consequently, the only energetically accessible electronic states for tunneling are the $p_z$ states of the graphite substrate.
As shown in our previous work~\cite{palinkas}, the $dI/dV$ spectra exhibit the characteristic "V"-shaped profile of graphite when the tip does not penetrate the molecular layer.
Likewise, the case discussed here, when the tip breaches the alkane layer, the $dI/dV$ curve retains the same "V"-shaped conductance, as illustrated by the red curve in Fig.~\ref{fig:1}e.
While measuring the spectra, the tip remains stationary, and the probability of forming or breaking conduction channels through the molecules is low.
As a result there is not much extra noise in the $dI/dV$ compared to the UHV cleaved case.
Furthermore, charge transport through the molecular layer contributes only a constant background to the spectrum.
However, the most striking feature of the $dI/dV$ measurements emerges when the clean surface is measured.
In this case, a clear gap is observed in the graphite tunneling conductance curve, as shown by the blue spectrum in Fig.~\ref{fig:1}e.
This gap is present in the tunneling conductance measurements of graphene~\cite{phonongap,Brar2010-fj,Natterer2015-xy} and graphite~\cite{Yin2020-ck} and is due to the strong suppression of elastic scattering near the Fermi level because of a momentum mismatch between the tunneling electrons and the large momentum of the graphene/graphite states near the $K$ points~\cite{Wehling2008}.
At larger energies, the momentum mismatch is overcome by scattering off large momentum phonons~\cite{Natterer2015-xy}.
The energies of these phonon modes (56.8 and 80.9 meV) are shown by dashed green lines in Fig.~\ref{fig:1}e, above which this inelastic tunneling pathway significantly increases the $dI/dV$ signal.
The strong conductance enhancement above the $\pm 80$ meV range results from the mixing of the electronic states at large momentum near $K$ and $M$ points~\cite{Natterer2015-xy} with the nearly free-electron state at $\Gamma$ by electron-phonon coupling~\cite{Wehling2008}.
Yin et al.~\cite{Yin2020-ck} demonstrated that the phonon-induced gap vanishes when hydrogen molecules are adsorbed in the tunneling junction.
This phenomenon arises because the molecule situated between the tip and the sample introduces additional tunneling channels, resulting in additional conductance pathways and therefore an enhanced $dI/dV$ signal in the phonon-gap region of the spectra.
Our results reinforce the finding that the phonon-induced gap is observable only in ultra-clean samples, where the tunneling junction is free from extraneous molecules.
In the case of n-alkane contamination the additional conductance channels due to the molecules smear out the phonon-gap, playing a similar role as the hydrogen molecules in the experiments of Yin et al.~\cite{Yin2020-ck}.

Apart from the disappearance of the phonon-gap (Fig.~\ref{fig:1}e), the $I(z)$ dependence is the strongest indicator that there is n-alkane contamination in the tunnel junction.
To understand the influence of the contaminant layer, let us consider first the textbook case of a clean tunnel junction.
Usually the tunneling current is calculated by evaluating the Bardeen tunneling matrix element~\cite{Bardeen1961-ft} in the Tersoff-Hamann approximation,~\cite{Tersoff1983-vf,Tersoff1985-kb} where the tip wave function has spherical symmetry.
We show here the expression derived by Feenstra~\cite{Feenstra2021-bf} at the tip apex position:

\begin{equation}
I \propto \int_0^{eV} \int_{\mathbf{k}_{\parallel}} d V d^2 \mathbf{k}_{\parallel} \, \left| \psi_{\mathbf{k}_{\parallel}} \right|^2 \left[ f_{\mathrm{t}}(E_{\mathbf{k}_{\parallel}}) - f_{\mathrm{s}}(E_{\mathbf{k}_{\parallel}}, V) \right]
\label{eq:current}
\end{equation}

The above equation (\ref{eq:current}) gives the contribution of a single band in the sample, labeled by the momenta parallel to the sample surface $\mathbf{k}_{\parallel}$, in the energy window defined by the Fermi functions $f_{\mathrm{s}}$ and $f_{\mathrm{t}}$ and the sample bias $V$.
This expression encompasses two important effects, one is that different bands will have a different decay of the sample wave function $\left| \psi_{\mathbf{k}_{\parallel}}(z) \right|^2$ into the vacuum and importantly that the decay can vary strongly with parallel momentum $\mathbf{k}_{\parallel}$~\cite{Zhang2008-hf,Zhang2015-sv,Feenstra2021-bf,Feenstra1987-zu}.
This latter effect is usually expressed as: $\left| \psi_{\mathbf{k}_{\parallel}} \right|^2 \propto \exp \left( -2 \kappa z \right)$, where

\begin{equation}
\kappa = \sqrt{\frac{2 m_0}{\hbar^2} \left( \bar{\phi} + \frac{eV}{2} \right) + \mathbf{k}_{\parallel}^2}
\label{eq:decay}    
\end{equation}

where $m_0$ is the free electron mass, $\hbar$ Planck's constant and $\bar{\phi}$ the average tunneling barrier height, usually taken as the average of the tip and sample work functions.
This approximation is generally valid for metallic surfaces, as demonstrated by the $I(z)$ spectrum of the Au (111) surface shown in Fig.~\ref{fig:1}f.
The measured decay constant $\kappa$ for the gold spectrum is $1.1 \pm 0.04$ \AA$^{-1}$, at a tip - sample bias of 0.5 V.
Just as a comparison, using the work functions of Au (111) (5.33 eV) and Pt (5.5 to 5.8 eV)~\cite{Derry2015-yz} and taking into account the above bias voltage, the calculated decay rate is $\kappa = 1.16$ to $1.18$ \AA$^{-1}$, in reasonably good agreement with the measured value.
We assume that tunneling predominantly occurs into states near the $\Gamma$ point, where $\mathbf{k}_{\parallel} = 0$, as these states exhibit the slowest decay into the vacuum.
By far the biggest uncertainty in this $\kappa$ value is the work function of the Pt/Ir STM tip, as the value for a clean Au (111) surface is well known~\cite{Derry2015-yz}.
Therefore it is worth estimating this based on the slope of the log of the $I(z)$ curve.
Considering the measured decay rate on gold of $1.1 \pm 0.04$ \AA$^{-1}$, the work function of the tip is estimated to be $4.5 \pm 0.7$ eV.
With this work function, the decay rate of an $I(z)$ curve on graphite, should be $1.06 \pm 0.04$ \AA$^{-1}$, based on a graphite work function of 4.6 eV~\cite{Takahashi1985-wp}, assuming a bias voltage of 0.5 V outside the phonon gap, such that tunneling occurs into states around $\Gamma$ due to coupling to large momentum phonon modes~\cite{phonongap,Wehling2008,Natterer2015-xy}.
Now, let's compare this expected $\kappa$ value to the one measured on the UHV cleaved and contaminated surfaces respectively.
In the case of the clean surface $I(z)$ shown in Fig.~\ref{fig:1}f, we calculate the decay rate to be $1.4 \pm 0.3$ \AA$^{-1}$, significantly higher than the expected value of $1.06 \pm 0.04$ \AA$^{-1}$.
This behavior is well understood: for materials with weak van der Waals coupling between layers, the STM tip can exert attractive forces on the sample that are comparable in magnitude to the interlayer binding forces~\cite{Georgi2017}.
This causes the top graphene layer to lift, which reduces the tip-sample distance beyond the expected decrease from the piezo motion.
Consequently, the tip-sample distance decreases faster than the assumed decrease based on the piezo $z$ displacement, leading to a faster rise in the current than what would be expected from the rigid tunneling model.
The behavior of the contaminated surface (Fig.~\ref{fig:1}f), on the other hand is more surprising.
For this particular $I(z)$ curve, the $\kappa$ we measure is $0.5 \pm 0.2$ \AA$^{-1}$.
This is anomalously small, and clearly beyond the simple tunneling model, because assuming equation~\ref{eq:decay} to hold, the work function of the sample would have to be negative.
We mention that we calculated the $\kappa$ values discussed above in the $z$ range of 0 to 1 \AA, where 0 is the initial position of the tip as determined by $I_{\mathrm{t}}$ and we retract the tip to 1 \AA\ distance.

To understand the unexpectedly low decay rate observed on the contaminated surface, we examine the $I(z)$ curves measured on this sample in more detail.
We mapped tens of thousands of $I(z)$ curves on homogeneous sample areas of $10 \times 10$ nm$^2$.
Each measurements begins by retracting the tip from a fixed tunneling current setpoint, then approaching again, followed by turning on the current feedback stabilization.
Starting at a setpoint of 80 pA, we increased the setpoint incrementally, thereby moving the tip closer to the sample surface.
The starting setpoint of $80$ pA was chosen because this is the setpoint at which the $I(z)$ curves contain enough data to evaluate the decay rate $\kappa$ and within this tunneling regime the tip starts to penetrate the n-alkane layer.
We plot the current values as a 2D histogram at specific tip - sample separations (see Fig.~\ref{fig:2}a-d), where the tip sample distance of 0 refers to the starting point of each curve, which was set by the current feedback setpoint.
For clarity we also plot as dashed lines the averaged $I(z)$ curves and a curve measured, at the same setpoint, on a clean (UHV cleaved) surface.
The measurements shown on Fig.~\ref{fig:2}a-d, confirm the trend displayed by the single spectra on Fig.~\ref{fig:1}f, namely that the current on the contaminated surface decays much more slowly compared to the clean surface.
Not only is the $\kappa$ of these spectra anomalously small, but in certain cases we still get a current in the range of a few 10 pA at $z = 7$ or $8$ \AA\ away from the surface.
This cannot be explained by pure tunneling, as the current should have decayed by over seven orders of magnitude at this $z$, as in the case of the clean surface. 

\begin{figure}[h!]
    \centering
    \includegraphics[width=\textwidth]{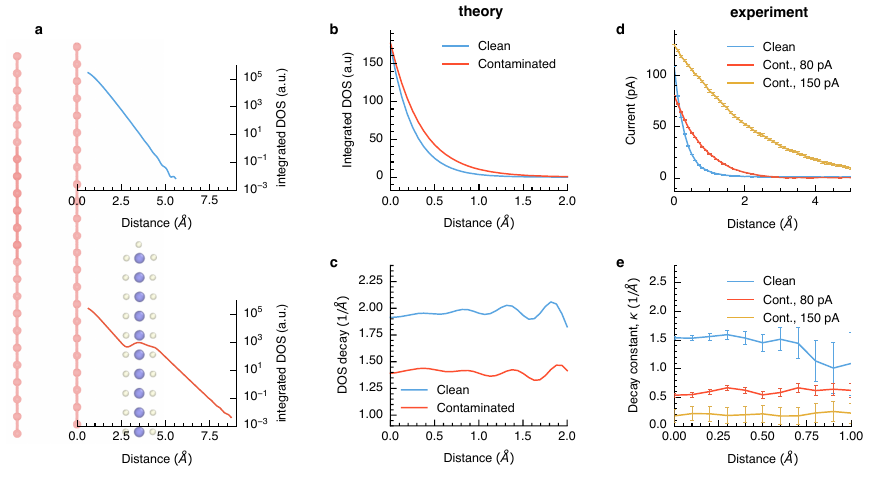}
    \caption{\textbf{\emph{Ab-initio} calculations and measured decay constants.}
    \textbf{(a)} DOS decay of a bilayer graphite surface with and without an alkane monolayer on top.
    The alkane layer modifies the decay of the graphite wave functions into the vacuum.
    The zero value of the distance coordinate is the position of the topmost C atom.
    The DOS values are integrated from the Fermi level to 0.5 eV for both cases.
    Oscillations of the DOS at large distances are due to numerical error.
    \textbf{(b)} Decay of the \emph{ab-initio} calculated DOS into the vacuum of the pure bilayer graphite surface and the n-alkane covered one.
    Both the clean and alkane covered DOS plot starts from the same DOS value to match STM $I(z)$ conditions.
    Same data as in (a).
    \textbf{(c)} Decay constant of the calculated DOS, obtained by differentiating the $\mathrm{log}$ of the DOS values versus distance from the surface $z$.
    \textbf{(d)} $I(z)$ curves measured on clean and contaminated graphite samples, with the tip at various distances inside the contaminant monolayer.
    Current setpoints shown in the legend.
    \textbf{(e)} Decay constants ($\kappa$, see eq.~\ref{eq:decay}) for the $I(z)$ curves in (d).
    The error bars correspond to two standard deviations of the current noise at zero current (tip far from sample).
    The decay constant and DOS decay are calculated by taking the $log$ of the values in panels b and d and calculating the derivative with respect to distance $z$.
    In panels d and e, the tip - sample distance defined by the current setpoint is chosen as $z = 0$.
    }
    \label{fig:4}
\end{figure}

In light of the anomalously low $\kappa$ value observed in our experiment, it is necessary to examine the mechanisms that can lead to a reduction in $\kappa$, or equivalently, in the tunneling barrier height.  
In general, for tunneling, the influence of the image potential must be taken into account.
When image charge effects are included, the barrier height and width are reduced, resulting in a smaller $\kappa$ value~\cite{Simmons1963,Akkerman2007,Huisman2009,Trouwborst2011}.  
This reduction can reach approximately 1 eV for metal–insulator–metal tunnel junctions~\cite{Akkerman2007}.
However, this effect cannot account for the $\kappa \approx 0.5$~\AA$^{-1}$ values observed in our measurements, as this would correspond to a barrier height $\bar{\phi}$ of 1.25 eV, according to Equation~\ref{eq:decay}.  
A further reduction in barrier height can arise from the orthogonalization of metal and molecular wave functions, known as the “pillow effect”~\cite{Vzquez2007}.
Additionally, charge transfer between the molecule and the surface can induce a dipole that lowers the work function.  
Both work-function-reducing effects are included in our \emph{ab initio} calculations of wave function decay.
In our case, the alkane molecules are bound through the van der Waals interaction, so the charge transfer between the adsorbates and the graphite substrate is very weak~\cite{Long2012}.  
To quantify the charge transfer, we performed \emph{ab initio} calculations for an AB-stacked two-layer graphite slab terminated by n-alkane molecules (C\textsubscript{17}H\textsubscript{36}).
We found that the net charge transfer for each molecule is $0.02$ electrons per alkane chain.
This small charge rearrangement does not lead to any significant change in the work function.
The pillow and dipole effects together can reduce the decay rate of the wave functions into the vacuum by 0.5~\AA$^{-1}$, as shown in Fig.~\ref{fig:4}c.
However, the decay rate measured in our $I(z)$ curves is a factor of three smaller than the calculated value.

The extremely low decay constant of the $I(z)$ curves and the presence of a significant current at distances above 4 \AA\ from the setpoint distance are proof that there is another possibility for electron transport apart from tunneling.
When the surface is covered with n-alkane contamination, this additional transport channel is conductance through the alkane molecules~\cite{transport}.
Taking as an example the data measured at $250$ pA setpoint (Fig.~\ref{fig:2}d), the current remains approximately $50$ pA at a tip–sample distance of $4$ \AA, whereas it decays to zero for the clean surface.
The measured conductance is $200\,\mathrm{pS}$, comparable to that of n-alkane molecules in break-junction experiments~\cite{transport}.
As the tip retracts from the surface in the presence of contaminants, the $I(z)$ measurement effectively performs a molecular break-junction experiment.
Similar to break-junctions~\cite{Gerhard2017-jw}, the curves in Fig.~\ref{fig:2}a–d exhibit hysteresis between the approach and retract curves.
Specifically, the current persists over a greater tip-sample separation as the tip is retracted.
This hysteresis is negligible when the tip is at its farthest distance from the surface ($I_{\mathrm{t}} = 80$ pA), but becomes increasingly pronounced as the current setpoint increases, positioning the tip closer to the sample at the start of the measurement.
We attribute this effect to the van der Waals adhesion of n-alkane molecules to the tip, which exerts a pulling force on the molecules.
Upon retraction, elastic deformation and progressive delamination of the molecules from the substrate by the tip~\cite{Ju2024-jd}, is expected to occur.
This mechanism results in a higher tunneling current during the retract curve compared to the approach.
A schematic illustration of this process is shown in Fig.~\ref{fig:2}e.
This effect is similar to how capillary neck formation is stronger when retracting an AFM tip from the sample surface under humid conditions~\cite{Weeks2005-sk,Zitzler2002-cr}, as opposed to approaching.
Of course, the alkane chains can also permanently adhere to the tip, making tips that were "dipped" into the alkane layer, prone to showing anomalous $I(z)$ curves and phonon-gap absent dI/dV spectra even on clean sample surfaces.
In this case replacing the tip is necessary or cleaning on gold is recommended, by repeatedly dipping the tip into the gold surface.

\begin{figure}[h!]
    \centering
    \includegraphics[width=1\textwidth]{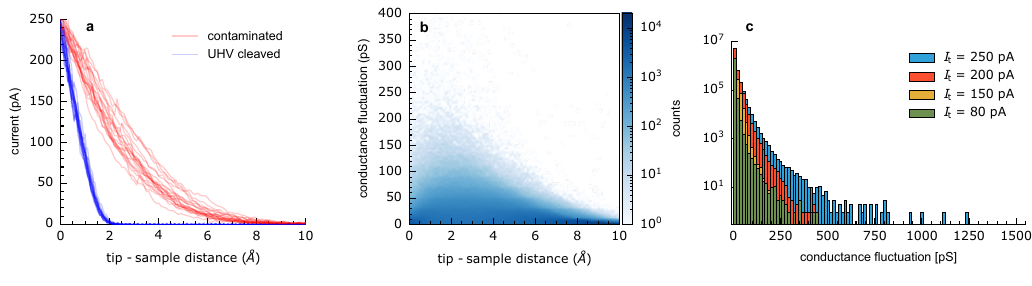}
    \caption{\textbf{Conductance fluctuations.}
    \textbf{(a)} Examples of individual $I(z)$ curves for the contaminated and clean graphite surfaces, measured by retracting the tip.
    \textbf{(b)} Histogram of conductance values for specific $I$ and $z$ values, for the alkane contaminated surface.
    Data from 20480 individual $I(z)$ curves.
    The spectra were measured by retracting the tip from the distance set by the current setpoint ($I_\mathrm{t}$ = 250 pA).
    The conductance fluctuation is calculated by subtracting the mean ($\left < I_{\mathrm{t}}(z) \right >$) from all individual $I(z)$ curves, dividing by the sample bias and taking the absolute value of it: $\left | \left ( I_{\mathrm{t}}(z) - \left < I_{\mathrm{t}}(z) \right > \right ) / V_{\mathrm{b}} \right |$.
    \textbf{(c)} Histogram of conductance fluctuations for all $z$ values.
    Increasing $I_{\mathrm{t}}$, we increase the number and magnitude of conductance fluctuations.
    In all cases $V_{\mathrm{b}}$ = 500 mV.
    The tip - sample distance defined by the current setpoint is chosen as $z = 0$.
    }
    \label{fig:3}
\end{figure}

During approach - retract cycles, stochastic formation and rupture of molecular junctions between the tip and sample result in pronounced current fluctuations.
The strong presence of these fluctuation with respect to the clean sample can be seen in Fig.~\ref{fig:3}a, where we compare 20 individual $I(z)$ spectra on the contaminated and clean graphite surfaces.
We define the conductance fluctuations at a specific tip - sample distance $z$ as: $\left | \left ( I_{\mathrm{t}}(z) - \left < I_{\mathrm{t}}(z) \right > \right ) / V_{\mathrm{b}} \right |$, where $\left < I_{\mathrm{t}}(z) \right >$ is the mean current value of all spectra in a particular measurement.
Plotting the conductance fluctuations versus tip - sample distance we see that on the contaminated surface, when we start the tip from deep within the alkane layer ($I_{\mathrm{t}} = 250 \text{ pA}$), the conductance fluctuations are on the order of the conductance associated with short (few-CH$_2$-unit) alkane chains, order of magnitude 10 to 100 pS (see Fig.~\ref{fig:3}b and Table~\ref{table1:label})~\cite{transport,Ju2024-jd}.
This suggests that as the tip continues to penetrate the alkane layer, the severing and reformation of tip--alkane--graphite junctions could contribute to the observed fluctuations.  
A similar dynamic effect was described by Kockmann et al.~\cite{Kockmann2009}.  
During the measurement, the tail of an alkane molecule can flip and make contact with the STM tip, temporarily bridging the tunnel gap.  
This interaction results in a sudden and significant increase in the tunneling current relative to the setpoint value.  
The noise arises from the dynamic nature of this contact, as the molecular tail intermittently reconnects with both the tip and the surface, leading to current fluctuations.
In Fig.~\ref{fig:3}c, we compare these fluctuations for various setpoints by plotting the conductance fluctuations at all $z$ values for tens of thousands of $I(z)$ curves, measured using increasing setpoint (derived from data shown in Fig.~\ref{fig:2}a-d).
We can observe that as $I_{\mathrm{t}}$ increases, both the magnitude and the number of fluctuations increases, as a result of the tip penetrating deeper into the contaminant layer.


We demonstrated that n-alkane contamination significantly alters the apparent $I(z)$ decay by enabling additional charge transport pathways through the molecular layer.
As a result, the intrinsic decay of the wave function on the alkane-covered surfaces is obscured.
Probing the intrinsic decay constant $\kappa$ on these surfaces, even at the lowest setpoint $I(z) = 80$pA is not possible.
Even under these conditions, the observed decay rate remains low, approximately 0.5~\AA$^{-1}$, indicating that charge transport through the alkane overlayer persists.
At setpoints lower than 80~pA, such as the 20~pA in Fig~\ref{fig:1}b, the $I(z)$ curves contain insufficient data to reliably evaluate $\kappa$.
This limitation, compounded by tip–surface van der Waals interactions that induce lifting on the clean graphite surface~\cite{Georgi2017}, hinders a direct comparison of the wave function decay above the alkane layer and the pristine surface.
To overcome these experimental constraints, we perform \emph{ab-initio} calculations using the Vienna Ab initio Simulation Package (VASP)~\cite{Kresse1999-tr} for an AB stacked two layer graphite slab and a surface terminated by n-alkane molecules.
For both systems, the DOS is calculated from electronic states near the $K$ points and integrated over the energy range extending from the graphite Fermi level to 0.5~eV, matching the experimental sample bias.
The calculated electron density as a function of distance from the surface is presented in Fig.~\ref{fig:4}a.
We can observe that the alkane layer perturbs the decay of the electronic states above the sample surface.
Since normal alkanes lack electronic states within the considered energy range, the overlayer acts as a perturbation to the graphite $p_{\mathrm{z}}$ states.
As a result, the decay observed in Fig.~\ref{fig:4}a is the attenuation of the graphite states through the molecular layer.
Consequently, the local charge density detected by STM above the alkanes primarily originates from graphite $p_{\mathrm{z}}$ orbitals.
At sample biases outside the phonon gap there is an additional contribution from electronic states near the $\Gamma$ point, mediated by electron–phonon coupling~\cite{phonongap,Wehling2008}.
The calculated DOS excludes contributions from $\Gamma$-point states, but it allows us to explore, how the alkane coverage changes the decay of the states around $K$.
In Fig.~\ref{fig:4}b, we plot the spatial decay of the DOS, starting from the same initial value.
The presence of the alkane layer leads to a slower decay of the graphite-derived states within the considered energy range.
The change in decay is even more evident if we calculate the logarithmic derivative of the DOS, according to $\mathrm{d} \ln(\mathrm{DOS}) / \mathrm{d}z$, as shown in Fig.~\ref{fig:4}c.
The decay rate of the clean surface is in excellent agreement with the $\kappa \approx 2$ measured by Zhang et al~\cite{phonongap} on graphene, inside the phonon gap.
The oscillations of the DOS decay at large surface distances are due to numerical error.
Compared to this, the calculated decay of the electronic states above the alkane layer is $\sim$30\%\ smaller, showing that the n-alkane overlayer is expected to lead to a decrease in the $\kappa$ measured above the molecules.
With a sufficiently low noise STM setup this could be measured in future experiments.
It should be noted that the calculated DOS decay is not equivalent to the current decay measured in STM, as it neglects the previously mentioned $\Gamma$-mediated tunneling channels, the tip electronic structure, and the tunneling matrix element.

Figure~\ref{fig:4}d,e summarizes the key observations from our $I(z)$ measurements, regarding the extracted current decay rates $\kappa$.
For the specific tips used in our measurements, the decay rate on the clean surface is typically in the range of $\sim$1.5~\AA$^{-1}$, only approaching the expected value of $\sim$1.1~\AA$^{-1}$ at large tip–sample separations, where the van der Waals attraction between the tip and top graphene layer is negligible~\cite{Georgi2017} (blue data in Fig.~\ref{fig:4}e).
Although, here the current noise is already quite large leading to a the larger error bars shown.
This behavior indicates that tip-induced lifting of the top graphene layers~\cite{Georgi2017} persists until the largest tip–sample distances.
The $I(z)$ curves recorded as the tip penetrates the n-alkane layer exhibit anomalously low $\kappa$ values, even for the smallest setpoint used in our measurements ($I_\mathrm{t} = 80$~pA) owing to charge transport through the alkane layer.

\section*{Conclusions} 
$I(z)$ measurements provide access to the local barrier height~\cite{Olesen1996-rd,Sharma2025-ok} and can also reveal signatures of electron–phonon coupling in graphene~\cite{phonongap}, among other properties.
Yet it is applied less often to van der Waals materials, partly because their current-decay behavior remains poorly understood. For such materials we observe that ambient n-alkane contamination~\cite{palinkas} causes anomalous current decay, consistent with charge transport along the alkane backbone of one or more molecular chains.
This additional molecular break-junction-like behavior introduces fluctuations in the tunneling current, which degrade the signal-to-noise ratio in atomic-resolution images and obscure the measurement of the intrinsic decay rate of the electronic wave function into the vacuum.
Using \emph{ab-initio} calculations, we were able to show that the alkane overlayer reduces the decay rate of the graphite wave functions into the vacuum, possibly enabling the unperturbed measurement~\cite{palinkas} of the molecular layer.

In tunneling spectroscopy measurements, charge transport through the contaminant layer obscures the phonon gap~\cite{phonongap} by introducing parallel conduction channels in addition to the vacuum tunnel gap~\cite{Yin2020-ck}.
The absence or presence of the phonon gap has often been attributed to tip-related effects.
While the structure and cleanliness of the tip~\cite{Yin2020-ck} can indeed influence its detection, we show that alkane surface contamination can suppress the phonon gap.
Finally, we mention that n-alkanes are also deposited onto samples prepared in glove-box environments, therefore proper removal of the surface contamination is necessary by vacuum annealing at a minimum of 200$^{\circ}$C, as we have shown previously~\cite{palinkas}.
We demonstrate that a simple $I(z)$ measurement serves as a practical diagnostic for surface contamination, since the current decay rate is always anomalously low compared to the expected value of $\kappa \approx 1.1$ \AA$^{-1}$, when n-alkane surface contamination is present.

\section*{Acknowledgement}

GK supported by the DKOP-23 doctoral excellence program and the ÚNKP-23-3 New National Excellence Program of the Ministry for Culture and Innovation from the source of the National Research, Development and Innovation Fund.
Funding from the "Lendület", Momentum program of the Hungarian Academy of Sciences through project LP2024-17 and from the National Research, Development, and Innovation Office (NKFIH) in Hungary, through the Grants K-146156, PD-146479, K-134258, FK-142985, TT-IPARI-KR-2022-00006, TKP2021-NKTA-05, KKP 138144 and the Excellence grant 151372 are acknowledged. MS and AP acknowledge the support from the János Bolyai Fellowship of the Hungarian Academy of Sciences.
MS acknowledges the Digital Government Development and Project Management Ltd. for awarding access to the Komondor HPC facility based in Hungary.

\subsection*{Data analysis and availability}
Data analysis of the STM measurements was carried out using the open-source Python tool: rhkpy \cite{rhkpy}.
Raw data files and the Python scripts used to create the figures in this manuscript are available at Zenodo, DOI: 10.5281/zenodo.17469441.


\section*{Competing interests}
The authors declare no competing interests.

\printbibliography

\end{document}